\documentclass[12pt]{elsart}

\usepackage{amsmath}
\usepackage{amsfonts}
\usepackage{graphicx}

\begin{document}

\begin{frontmatter}

\title{Statistical measures applied to metal clusters: \\ evidence of magic numbers}

\author[jsr]{Jaime Sa\~{n}udo}
\ead{jsr@unex.es} and
\author[rlr]{Ricardo L\'{o}pez-Ruiz}
\ead{rilopez@unizar.es}

\address[jsr]{
Departamento de F\'isica, Facultad de Ciencias, \\
Universidad de Extremadura, E-06071 Badajoz, Spain, \\
and BIFI, Universidad de Zaragoza, E-50009 Zaragoza, Spain}

\address[rlr]{
DIIS and BIFI, Facultad de Ciencias, \\
Universidad de Zaragoza, E-50009 Zaragoza, Spain}


\begin{abstract}
In this work, a shell model for metal clusters up to 220 valence electrons 
is used to obtain the fractional occupation  probabilities of the electronic orbitals.
Then, the calculation of a statistical measure of complexity 
and the Fisher-Shannon information is carried out. 
An increase of both magnitudes with the number of valence electrons is observed.
The shell structure is reflected by the behavior of the statistical complexity.
The magic numbers are indicated by the Fisher-Shannon information. 
So, as in the case of atomic nuclei, the study of statistical indicators 
also unveil the existence of magic numbers in metal clusters.
\end{abstract}

\begin{keyword}
Statistical Indicators; Metal Clusters; Shell Structure; Magic Numbers
\PACS{36.40.-c, 89.75.Fb.}
\end{keyword}

\end{frontmatter}

\maketitle

Nowadays the calculation of information-theoretic measures on quantum systems 
has received a special attention \cite{gadre1987,panos2005,sanudo2008-}.
The application of these indicators to atoms or nuclei reveals some properties 
of the hierarchical organization of these many-body systems \cite{panos2009,sanudo2009}.  
In particular, entropic products such as Fisher-Shannon information and
statistical complexity present two main characteristics when applied
to the former systems. On one hand, they display an increasing trend 
with the number of particles, electrons or nucleons. 
On the other hand, they take extremal values on the closure of shells.
Moreover, in the case of nuclei, the trace of magic numbers is displayed 
by these statistical magnitudes \cite{lopezruiz2009}.

Metal clusters are useful quantum systems to understand how the physical 
properties evolve in the transition from atom to molecule to small particle
to bulk solid \cite{meiwes2000,broglia2004}. They also present a shell structure 
where it can applied the statistical indicators before mentioned. As in the case of
atoms and nuclei, the fractional occupation probabilities of valence electrons
in the different orbitals can capture the shell structure. This set of probabilities can be used
to evaluate the statistical quantifiers for metallic clusters as a function of the number of valence
electrons. Similar calculations have been reported for the electronic atomic 
structure \cite{panos2009,sanudo2009} and for nuclei \cite{lopezruiz2009}.
In this work, by following this method, we undertake the calculation of statistical
complexity and Fisher-Shannon information for metal clusters. 

The jellium model provides an accurate description of some simple metal clusters.
In this model, a valence electron is assumed to interact with the average potential
generated by the other electrons and the ions \cite{meiwes2000,broglia2004}.
The confinement potential in the Schr\"odinger equation leading to shell structure 
is taken as a potential intermediate between the three-dimensional harmonic oscillator 
and the three-dimensional square well. This yields a filling of shells with a number $N$
of valence electrons given by the series: 2, 8, 18, 20, 34, 40, 58, 68, 70, 92, 106,
112, 138, 156, 166, 168, 198, 220 and so on.
Each shell is given by $(nl)^w$, 
where $l$ denotes the orbital angular momentum ($l=0,1,2,\ldots$), 
$n$ counts the number of levels with that $l$ value,
and $w$ is the number of valence electrons in the shell, $0\leq w\leq 2(2l+1)$. 

As an example, we explicitly give the shell configuration of a metal cluster formed
by $N=58$ valence electrons. It is obtained:
\begin{equation}
 (N=58):  (1s)^2(1p)^6(1d)^{10}(2s)^2(1f)^{14}(2p)^6(1g)^{18}\,.
\end{equation}
The fractional occupation probability distribution of electron 
orbitals $\{p_k\}$, $k = 1,2,\ldots,\Pi$,  being $\Pi$ the 
number of shells, can be defined in the same way 
as it has been done for calculations in atoms and nuclei \cite{panos2009,sanudo2009,lopezruiz2009}. 
This normalized probability distribution $\{p_k\}$ $(\sum p_k=1)$ is easily found by dividing 
the superscripts $w$ by the total number $N$ of electrons. 
Then, from this probability distribution, the different statistical magnitudes 
(Shannon entropy, disequilibrium, Fisher information, statistical complexity and
Fisher-Shannon entropy) can be obtained.

Here, we undertake the calculation of entropic products, a statistical measure of 
complexity $C$ and the Fisher-Shannon entropy $P$, that result from the product
of two statistical quantities, one of them representing the information content of the system,
and the other one giving an idea of how far the system is from the equilibrium.
The classical indicator of information is the Shannon entropy, that in the discrete
version is expressed as
\begin{equation}
S  =  -\sum_{k=1}^{\Pi}p_k\log p_k \,. 
\end{equation}
Any monotonous function of $S$ can also be used for this purpose, such as  
the exponential Shannon entropy \cite{dembo1991},
$H = e^{S}$ or $J = {1\over 2\pi e}\; e^{2S/3}$.
The indicator of how much concentrated is the probability 
distribution of the system can be related with some kind of distance to the
equilibrium distribution, that in our case is the equiprobability.
The disequilibrium $D$ given by
\begin{equation}
D  = \sum_{k=1}^{\Pi}(p_k-1/\Pi)^2 \,,
\end{equation}
and the Fisher information $I$ by
\begin{equation}
I = \sum_{k=1}^{\Pi} {(p_{k+1}-p_k)^2\over p_k}\;,
\end{equation}
where $p_{\Pi+1}=0$, are two useful parameters in this direction.
Then, the statistical measure of complexity, $C$, 
the so-called LMC complexity \cite{lopez1995,lopez2002}, is defined as 
\begin{equation}
C = H\cdot D\;,
\end{equation}
and the Fisher-Shannon information \cite{vignat2003,romera2004,szabo2008}, $P$, 
is given by
\begin{equation}
P = J\cdot I \;.
\label{eq-p}
\end{equation}

\begin{figure} 
\centerline{\includegraphics[width=12cm]{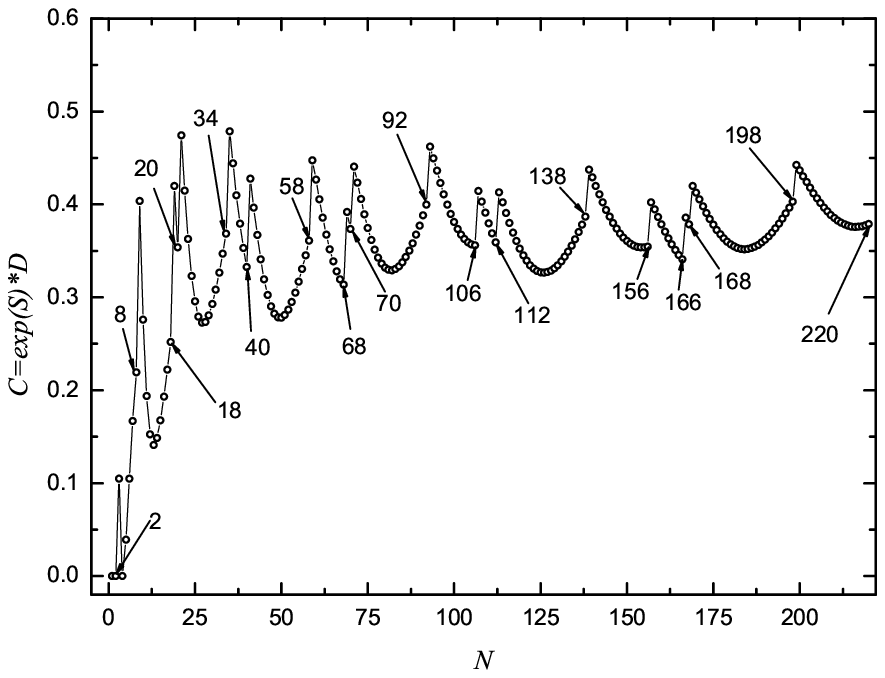}}  
\caption{Statistical complexity, $C$, vs. number of valence electrons, $N$.
The arrows indicate the positions of closed shells.}  
\label{fig1}  
\end{figure}

The statistical complexity, $C$, of metal clusters as a function of the number of 
valence electrons, $N$, is given in Fig. \ref{fig1}. 
We can observe in this figure that this magnitude fluctuates around an slightly increasing 
average value $N$. This trend is also found for the electronic structure of atoms \cite{sanudo2009}
and for the shell structure of nuclei \cite{lopezruiz2009},
reinforcing the idea that in general complexity increases with the number of units forming a system.
However, the shell model supposes that the system encounters certain ordered rearrangements 
for some specific number of units (electrons or nucleons) that coincide with closed shells.
In the present case, this fact is reflected by the notable increase of $C$
in the metal clusters with one valence electron more than those with closed shells, 
which are indicated in Fig. \ref{fig1}, just as happens 
for atoms when one electron is added to noble gases or when one nucleon is added to 
a closed shell in nuclei. Observe that some major shells do not show local minima 
at their closing. This effect is due to the number of valence electrons belonging to each shell:
a shell with a few valence electrons displays a local minimum of $C$ when is closed, 
but this is not the case when the number of valence electrons in a shell increases.

\begin{figure} 
\centerline{\includegraphics[width=12cm]{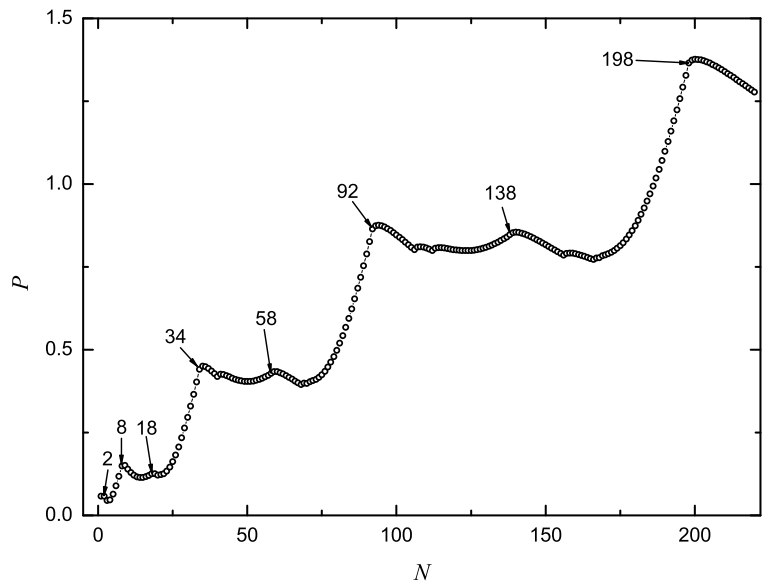}}     
\caption{Fisher-Shannon entropy, $P$, vs. the number of valence electrons, $N$.
The arrows indicate prominent closed shells that are magic numbers.
For details, see the text.}  
\label{fig2}  
\end{figure}

\begin{figure}  
\centerline{\includegraphics[width=12cm]{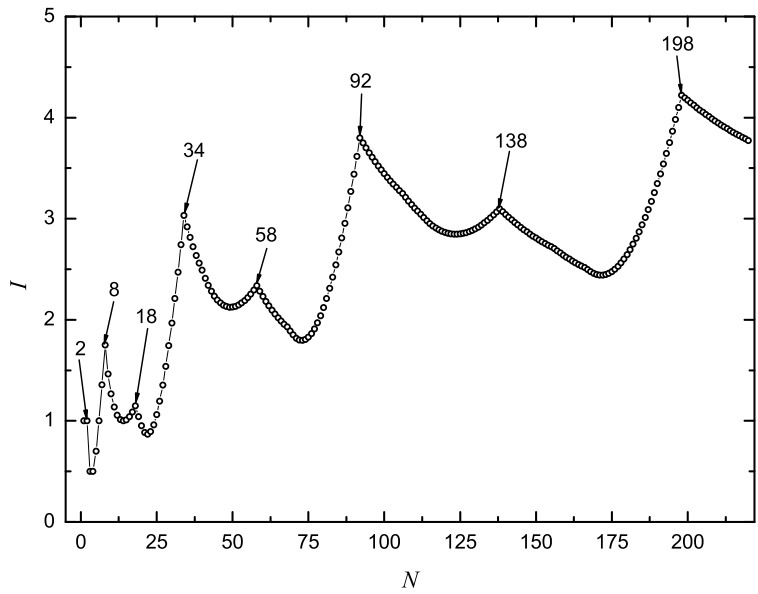}}     
\caption{Fisher information, $I$, vs. the number of valence electrons, $N$.
The arrows indicate prominent closed shells that are magic numbers.}  
\label{fig3}  
\end{figure}

The Fisher-Shannon entropy, $P$, of metal clusters as a function of $N$ is given in Fig. \ref{fig2}. 
It presents an increasing trend with $N$. The spiky behavior of $C$ provoked by 
the shell structure is still present for $P$ but becomes smoother in this case.
$P$ displays notable peaks only at a few $N$ related with the filling of some major shells, 
concretely at the numbers $2,8,18,34,58,92,138,198$. It must be remarked that, similarly as happens
with $C$, the maximum values of $P$ are taken on the nuclei with one unit more than the former series,
although now the difference is slightly appreciable. 
Only peaks at $20$ and $40$ disagree with the sequence of magic numbers
\{2, 8, 18, 20, 34, 40, 58, 92, 138, 198\} obtained from experimental data, 
for instance, for $Na$ clusters \cite{martin1990} and for $Cs$ clusters \cite{pedersen1990,lange1990}.
Let us observe that the magic numbers are basically marked by the Fisher information such as
it can be seen in Fig. \ref{fig3}.

In summary, the behavior of the statistical complexity $C$ and the Fisher-Shannon
information $P$ with the number of valence electrons in metal clusters has been reported.
The increasing trend of these magnitudes with the 
number of valence electrons, $N$, has been found. The method that uses the fractional 
occupation probabilities has been applied to calculate these statistical indicators.
As in the case of atoms and nuclei, the shell structure is well displayed by the spiky 
behavior of $C$. On the other hand,
$P$ shows an smoother behavior but with relevant peaks just on the major 
shells that coincide with the series of magic numbers in metal clusters.
Therefore, the qualitative study of metal clusters 
by means of statistical indicators unveil certain physical properties of them.
In fact, we can conclude that this type of statistical measures is able to enlighten 
some conformational aspects of quantum many-body systems.

\section*{Acknowledgements}
 The authors acknowledge some financial support from  the Spanish project
 DGICYT-FIS2009-13364-C02-01. J.S. also thanks to the Consejer\'ia de 
 Econom\'ia, Comercio e Innovaci\'on of the Junta de Extremadura (Spain) for 
 financial support, Project Ref. GRU09011.

\newpage  
  
\end{document}